\documentclass[pdflatex,sn-mathphys-num]{sn-jnl}

\usepackage{amsmath,amssymb}
\usepackage{bbold}
\usepackage{float}
\usepackage{graphicx}
\usepackage{dcolumn}
\usepackage{bm}
\usepackage{hhline}
\usepackage{color}
\usepackage[usenames,dvipsnames]{xcolor}

\newcommand{\onlinecite}[1]{\hspace{-1 ex} \nocite{#1}\citenum{#1}}

\begin{document}

\title{Phase-sensitive cascade quantum amplifier with nearly noiseless operation}

\author*[1]{\fnm{Ilari} \sur{Lilja}}\email{ilari.lilja@aalto.fi}

\author*[1]{\fnm{Ekaterina} \sur{Mukhanova}}\email{ekaterina.mukhanova@aalto.fi} 

\author[1]{\fnm{Michael} \sur{Perelshtein}} 

\author[1]{\fnm{Kirill} \sur{Petrovnin}}

\author[1]{\fnm{Stanislav} \sur{Khaldeev}}

\author[2]{\fnm{Visa} \sur{Vesterinen}}

\author[1]{\fnm{Gheorghe-Sorin} \sur{Paraoanu}}

\author[1]{\fnm{Pertti} \sur{Hakonen}} 

\affil[1]{\orgdiv{QTF Centre of Excellence, Department of Applied Physics}, \orgname{Aalto University}, \orgaddress{\street{Otakaari 1}, \city{Espoo}, \postcode{02150}, \country{Finland}}}

\affil[2]{\orgdiv{QTF Centre of Excellence}, \orgname{VTT Technical Research Centre of Finland Ltd}, \orgaddress{\street{Tekniikantie 21}, \city{Espoo}, \postcode{02150}, \country{Finland}}}





\abstract{
Phase-sensitive parametric devices enable quadrature-selective amplification with the potential for sub-quantum-limited noise performance. 
In this work, we investigate the operation of a SQUID-based Josephson Parametric Amplifier (JPA), comparing its performance in the phase-preserving and phase-sensitive regimes. The device, fabricated using VTT SWAPS technology, is driven in a three-wave mixing configuration and characterized in a reflection-based measurement setup at millikelvin temperatures. To directly probe the noise performance at low JPA gains, we employ a cascaded amplification scheme in which a Traveling-Wave Parametric Amplifier (TWPA) provides low-noise pre-amplification of the JPA output. 
In a phase-preserving operation, the JPA exhibits near-quantum-limited performance with a system noise temperature of $351\pm53$ mK at 6 GHz. In contrast, phase-sensitive operation yields a minimum system noise temperature of $94\pm12$ mK, well below the standard quantum limit of 288 mK. Our results demonstrate that a JPA–TWPA amplifier cascade opens the door to direct, high-fidelity probing of quantum devices without the need for background noise subtraction.}

\maketitle

\subsection*{Background}

Parametric amplifiers can be operated in two distinct regimes: either as phase-preserving (nondegenerate) or as phase-sensitive (degenerate) amplifiers \cite{caves1982quantum, Yurke1984, stenholm1986theory, clerk_introduction_2010, roy2016introduction}. In phase-preserving amplification, the signal frequency relative to the driving pump frequency is  $\omega_{\textrm{s}} \neq \omega_{\textrm{p}}/2$. As a consequence of energy conservation, an idler tone is generated such that $\omega_{\textrm{s}}+\omega_{\textrm{i}}=\omega_{\textrm{p}}$. Phase-preserving (PP) amplification is the most common operation mode of parametric amplifiers, and this type of operation adds, at minimum, an amount of noise equal to half a photon \cite{caves1982quantum,bergeal_phase-preserving_2010,caves2012quantum}. 

The quantum limit on the added noise of a linear amplifier follows from the requirement that the amplified field obeys the canonical bosonic commutation relations. If an input field operator, $\hat{a}_\textrm{in}$, is amplified with power gain $G$, the output field operator, $\hat{a}_\textrm{out}$, cannot simply be given by $G\hat{a}_\textrm{in}$, as this transformation does not preserve the commutation relation [$\hat{a},\hat{a}^{\dagger}$]= 1 \cite{caves1982quantum, clerk_introduction_2010}. To satisfy the commutation relations, an additional noise operator, $\hat{L}^{\dagger}$, must be introduced, yielding the correct input-output relation. For random noise $\langle \hat{L} \rangle = 0$, which leads to $[\hat{L}, \hat{L}^{\dagger}] = G - 1$ and to an additional term in the output field: $\hat{a}_{\textrm{out}} = \sqrt{G}\hat{a}_{\textrm{in}} + \hat{L}^{\dagger}$. Then, the commutation relation becomes:

\begin{equation}
    [\hat{a}_{\textrm{out}}, \hat{a}^{\dagger}_{\textrm{out}}] = [\sqrt{G}\hat{a}_{\textrm{in}} + \hat{L}^{\dagger}, \sqrt{G}\hat{a}_{\textrm{in}}^{\dagger} + \hat{L}]
    = G - (G - 1) = 1.
    \label{EQ:OutCommutator}
\end{equation}

The corresponding output fluctuations are obtained from the variance of the output field
\begin{equation}
\Delta\hat{a}_\textrm{out}^2=\left<\hat{a}_\textrm{out}^{\dagger}\hat{a}_\textrm{out}\right>-|\left<\hat{a}_\textrm{out}\right>|^2,
\end{equation} which, assuming that the amplifier noise is uncorrelated with the input field, becomes 
variance of the output field
\begin{equation}
\Delta\hat{a}_\textrm{out}^2=G\Delta\hat{a}_\textrm{in}^2+\left<\hat{L}\hat{L}^{\dagger}\right>.
\end{equation}
The second term represents the noise added by the amplifier and is commonly expressed as an equivalent input-referred noise temperature $\textrm{T}_\textrm{N}$, or equivalently as an added noise photon number.

In parametric amplifiers, different frequencies become coupled through their interaction with a nonlinear medium. As a result, noise originating at one frequency can reappear at a different frequency, linked via the conversion process. In the case of parametric amplification, this means that when measuring the signal's noise, one would observe not only its own amplified noise, but also an additional contribution from the corresponding idler. In frequency space, we denote the input signal and idler fields by $\hat{a}(\omega_s)$ and $\hat{a}(\omega_i)$, respectively. To determine the added noise of the JPA, commonly expressed as the amplifier noise temperature, we adopt the approach of Ref. \onlinecite{DPa_kipa_Grimsmo2022}. From Eq.~\ref{EQ:OutCommutator}, we obtain the following constraint for the output field:

\begin{equation}
\hat{a}_{\textrm{out}}(\omega_s)=\sqrt{G_\textrm{s}}\hat{a}_{\textrm{in}}(\omega_s)+\sqrt{G_\textrm{i}}\hat{a}_{\textrm{in}}^{\dagger}(\omega_i)+\sqrt{\frac{\gamma}{\kappa}}\left[(\sqrt{G_\textrm{s}}+1)\hat{b}_{\textrm{in}}(\omega_i)+\sqrt{G_\textrm{i}}\hat{b}_{\textrm{in}}^{\dagger}(\omega_i)\right],
\end{equation}

where $G_{s,i}$ is the power gain at signal and idler modes, and the third term on the right-hand side arises from the intrinsic cavity loss mode $\hat{b}_\textrm{in}(\omega_ \textrm{s})$ of the JPA that follows the usual commutation rules $[\hat{b}(\omega_s),\hat{b}^{\dagger}(\omega_i)]=\delta_{si}$, and $\gamma$ and $\kappa$ denote the intrinsic loss rate and external coupling rate of the cavity, respectively. 

%

%

The noise performance of the amplifier is characterized using the field quadratures $\hat{X}$ and $\hat{P}$. At the signal frequency $\omega_{\mathrm{s}}$, the quadrature operators are defined as $\hat{X}(\omega_\textrm{s})=(\hat{a}(\omega_\textrm{s})+\hat{a}^{\dagger}(\omega_\textrm{s}))/2$ and $\hat{P}(\omega_\textrm{s})=(\hat{a}(\omega_\textrm{s})-\hat{a}^{\dagger}(\omega_\textrm{s}))/(2i)$ \cite{agarwal_quantum_2012,DPA_Blais_2017}. The variance of the output quadrature, $\Delta{}X^2_{\textrm{out}}(\omega_\textrm{s})=\left<X_{\textrm{out}}(\omega_\textrm{s})^2\right>-\left<X_{\textrm{out}}(\omega_\textrm{s})\right>^2$, quantifies the output field fluctuations. In the limit of large signal and idler gains, $G_{\mathrm{s}},G_{\mathrm{i}}\gg1$, these fluctuations can be related directly to the input field variances at the signal and idler frequencies:

\begin{equation}
\left<\Delta{}X_{\textrm{out,PP}}^2(\omega_s)\right> = \left(G_\textrm{s}+\frac{\gamma}{\kappa}(\sqrt{G_\textrm{s}}+1)^2)\right)\left(\frac{n_\textrm{th}}{2}+\frac{1}{4}\right)+G_\textrm{i}\left(1+\frac{\gamma}{\kappa}\right)\left(\frac{n_\textrm{th}}{2}+\frac{1}{4}\right),
\label{Quadraqture_noise}
\end{equation}

where $n_{\mathrm{th}}=\left(e^{\hbar\omega/k_{\mathrm{B}}T_{\mathrm{th}}}-1\right)^{-1}$ is the thermal photon occupation and related to the physical temperature of the system, $T_{\mathrm{th}}$. We assume equal gain for the signal and idler modes, as is required by the symmetry of the conversion process. Referring all fluctuations to the JPA input, we can define the added noise at the signal mode of the JPA in the phase-preserving regime:

\begin{equation}
\underbrace{\frac{\left<\Delta{}X_{\textrm{out,PP}}^2(\omega_s)\right>}{G_s} \quad -\underbrace{\frac{1}{4}}_{\textrm{Vac (s)}}}_{\textrm{Added (s)}} = \quad 
\underbrace{\frac{1}{4}\left(1-\frac{1}{G_\textrm{S}}\right)}_{\textrm{Vac (i)}}
+\underbrace{n_\textrm{th}\left(1-\frac{1}{2G_\textrm{S}}\right)}_{\textrm{Thermal (s+i)}}
+\underbrace{\frac{\gamma}{\kappa}\left(n_\textrm{th}+\frac{1}{2}\right)}_{\textrm{Loss (s+i)}}
\end{equation}
where the symbol inside the parentheses signifies whether the source of noise was signal (s), idler (i), or both (s+i). The second term on the left-hand side represents the unavoidable quantum vacuum fluctuations at the signal quadrature input and is therefore not regarded as added noise by the amplifier. The added noise in this quadrature is consequently given by the right-hand side. The first term arises from the mixing of idler vacuum fluctuations into the signal mode. The second term describes the noise contribution from the finite physical temperature of the amplifier, also mixing from the idler mode in addition to the signal mode. The final term accounts for noise originating from internal cavity losses in the JPA, again for both signal and idler.

The corresponding expression for the orthogonal quadrature is identical, consequently, both quadratures contribute equally to the total output fluctuations. In the limit $n_{\mathrm{th}},\gamma\rightarrow0$, the input-referred added fluctuations in each quadrature are $1/4$, originating from the idler mode. Including the unavoidable zero-point fluctuations at the input, the total variance in each quadrature becomes $1/2$, corresponding to the vacuum zero-point energies $\hbar\omega_{\mathrm{s}}/2$ and $\hbar\omega_{\mathrm{i}}/2$ for the signal and idler modes, respectively. Since cavity-based Josephson parametric amplifiers (JPAs) are narrowband devices with $\omega_{\mathrm{i}}\approx\omega_{\mathrm{s}}$, the minimum noise of a coherent signal under \emph{phase-preserving} amplification, the standard quantum limit (SQL), is therefore $\hbar\omega_{\mathrm{s}}$ \cite{yurke_observation_1988}. One half of this quantum-limited noise is carried by the input signal itself, while the other half is added by the amplifier through coupling to the idler mode.

In \emph{phase-sensitive} (PS) parametric amplification, the signal and idler are degenerate, with $\omega_{\mathrm{s}}=\omega_{\mathrm{p}}/2$. The output field is now written in terms of relative signal and pump phase $\theta$ as:

\begin{equation}
\hat{a}_{\textrm{out}}=\sqrt{G_\textrm{s}}e^{-i\theta}\hat{a}_{\textrm{in}}+\sqrt{G_\textrm{s}}e^{i\theta}\hat{a}_{\textrm{in}}^{\dagger}+\sqrt{\frac{\gamma}{\kappa}}\left[(\sqrt{G_\textrm{s}}+1)e^{-i\theta}\hat{b}_{\textrm{in}}+\sqrt{G_\textrm{s}}e^{i\theta}\hat{b}_{\textrm{in}}^{\dagger}\right].
\end{equation}

Unlike phase-preserving amplification, the gain in phase-sensitive operation depends on the relative phase between the signal and the pump, allowing one field quadrature to be amplified beyond the phase-preserving gain while the conjugate quadrature is simultaneously deamplified. Because no independent idler mode is required, the transformation can satisfy the bosonic commutation relations without introducing the additional vacuum fluctuations that are inevitable in phase-preserving amplification \cite{yamamoto2003noise, malnou2018optimal}. The output quadrature variances are therefore
\begin{equation}
    \Delta\hat{X}_{\textrm{out,PS}}^2=G\Delta\hat{X}_{\textrm{in,PS}}^2, \ \ \ \ \Delta\hat{P}_{\textrm{out,PS}}^2=\Delta\hat{P}_{\textrm{in,PS}}^2/G,
\end{equation}
where $\hat{X}$ and $\hat{P}$ denote the amplified and deamplified quadratures, respectively. The reciprocal scaling of the two variances preserves the uncertainty product and produces quadrature squeezing.


For the amplified field quadrature, one obtains
\begin{equation}
\left<\Delta{}X_{\textrm{out,PS}}^2\right> = G_\textrm{PS}\left(\frac{n_\textrm{th}}{2}+\frac{1}{4}\right)+\frac{\gamma}{\kappa}(G_\textrm{PS}+1)\left(\frac{n_\textrm{th}}{2}+\frac{1}{4}\right),
\end{equation}
where $G_\textrm{PS} = 4G_{\mathrm{s}}$, with the factor of four arising from the coherent summation of the signal and idler quadratures. As in the phase-preserving case, the input-referred noise in the amplified quadrature can be written as
\begin{equation}
\underbrace{\frac{\left<\Delta{}X_{\textrm{out,PS}}^2\right>}{G_\textrm{PS}} \quad -\underbrace{\frac{1}{4}}_{\textrm{Vac (s)}}}_{\textrm{Added (s)}} = \quad
\underbrace{\frac{n_\textrm{th}}{2}}_{\textrm{Thermal (s)}}
+\underbrace{\frac{\gamma}{\kappa}\left(\frac{n_\textrm{th}}{2}+\frac{1}{4}\right)}_{\textrm{Loss (s)}}.
\end{equation}
%
%
Since the signal and idler are degenerate in the phase-sensitive configuration, there is no independent idler mode to contribute the additional $1/4$ quantum of input-referred noise. The added noise therefore arises solely from the finite physical temperature of the amplifier and from internal cavity losses in the JPA, the latter being reduced by a factor of two compared to the phase-preserving case. Since the orthogonal quadrature is deamplified, only the amplified quadrature contributes to the measured fluctuations. In the limit $n_\textrm{th},\gamma\rightarrow 0$, the amplifier therefore adds no input-referred noise, and the remaining fluctuations are solely those of the input vacuum. These correspond to the zero-point fluctuations of a single quadrature, with an energy of $\hbar\omega_{\mathrm{s}}/4$.

%
%
By constructing an experiment where the desired information is coded in only one of the quadratures, one can obtain nearly noiseless amplification and bypass the standard quantum limit \cite{salmanogli2025technical}. A detailed description of the operation of phase-sensitive three-wave mixing (3WM) can be found, e.g., in Refs. \cite{DPA_Blais_2017,DPa_kipa_Grimsmo2022}.

\subsection*{Experimental Techniques}

\begin{figure}[h]
    \noindent\centering{
    \includegraphics[width=120mm]{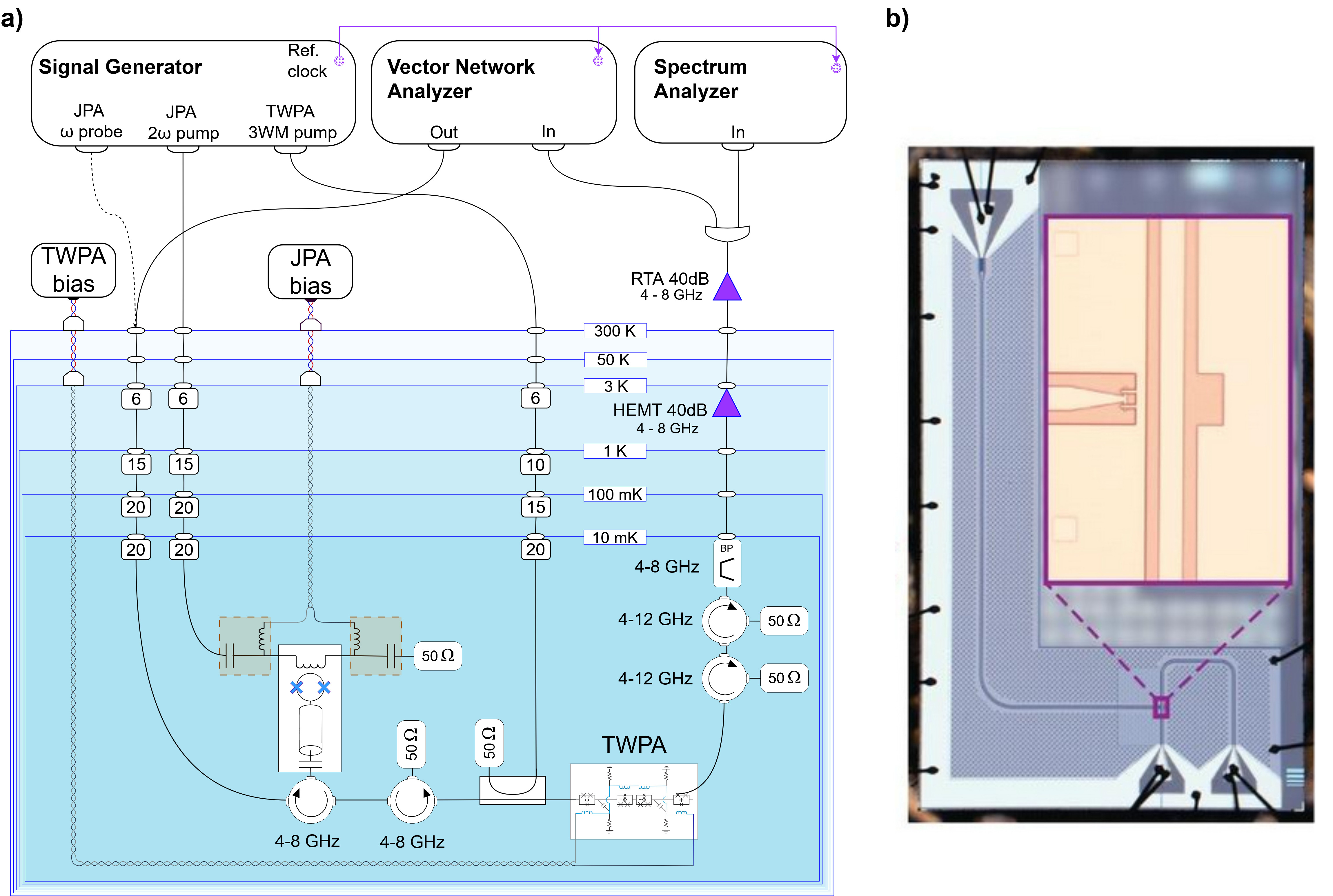}
    }
    \caption{
    \textbf{a)} Schematic of the sample and the low-temperature configuration of the experimental setup. For studies of phase-sensitive amplification, a TWPA is added to the output side of the JPA, where it acts as a pre-amplifier, with a circulator and a directional coupler/diplexer in between. This is done to ensure that the strong pump tone can be applied at the input of the TWPA without affecting the operation of the JPA.  \textbf{b)} An optical image of the sample with a close-up of the SQUID and its flux bias line (vertical strip). The JPA reflection port is seen at the top while the two pads at the bottom are for supplying flux bias and drive for the SQUID. Each pad is bonded using two Al wires.}
    \label{Setup}
\end{figure}

For direct demonstration of operation below SQL, we have employed a combination of a Josephson Parametric Amplifier (JPA) and a Traveling Wave Parametric Amplifier (TWPA) in cascade for phase-sensitive parametric amplification. Our JPA is fabricated with a Superconducting Quantum Interference Device (SQUID) providing the non-linear element. The JPA is manufactured by VTT using SWAPS technology, first discussed in Ref. \cite{gronberg2017}. The JPA consists of a quarter-wave transmission line resonator terminated by the SQUID to the ground with an integrated on-chip magnetic flux line for driving the SQUID loop with a strong pump tone \cite{yamamoto_flux-driven_2008,elo_broadband_2019}. The SQUID is driven in the 3WM regime \cite{renger2021beyond}, where a single pump photon is converted to the signal and idler photons. As such, the pump tone is applied at approximately twice the cavity resonance frequency. This places the pump far outside the measurement band, eliminating pump–signal overlap and rendering phase control easy. However, 3WM does require the SQUID to be biased by a constant DC magnetic field, which is implemented using a bias-T in the flux bias line outside the sample box. The JPA is operated in a reflection-based measurement setting, where the in- and outgoing fields are separated by a circulator. The device and the low-temperature configuration are illustrated in Fig. \ref{Setup}. Details on our cavity-based JPAs are found in Refs. \cite{Petrovnin2022,Petrovnin2024}.

The noise performance of the amplifier chain can be estimated with the Friis formula. This formula gives the total added noise for a cascade of amplifiers \cite{macklin2015near}:
\begin{equation}
    T_S = T_Q + T_{N,1}+\frac{T_{N,2}}{G_1}+\frac{T_{N,3}}{G_1G_2}+...,
    \label{Eq:Systemnoise}
\end{equation}
where the index indicates the order in which the amplifier appears in the amplification chain and $T_{\textrm{Q}}=\hbar\omega_s/2=144$ mK is the quantum contribution at $\simeq 6$GHz. The quantity $T_{\textrm{S}}$ is the system noise temperature, which includes noise contributions from all sources as well as the perceived increase in noise due to losses. The Friis formula shows that the added noise of the first amplifier in the amplifier chain is practically always the most important contributing factor to system noise, as all other amplifier noise contributions are suppressed by the gain factor of the preceding amplifiers.

For calibrating the system noise temperature $T_{\textrm{S}}$, we heated the mixing chamber of the cryostat and utilized the standard Y-factor method to set the reference point for system gain and noise temperature. This gives an approximate calibration, as the system as a whole is matched to 50 Ohms at multiple locations, and changes in the thermal gradients of rf-cables may slightly change thermal noise contributions coming from their losses. Consequently, we estimate an uncertainty of $\pm1$ dB due to various sources in our calibration of the cooled HEMT system noise temperature $T_{\textrm{S}}$. \footnote{The Y-factor calibration parameters were a 500 mK cold load, a 1 K hot load, a 1 MHz measurement bandwidth, and 100 averages. The calibration yielded $\sim{}10$ K for the LNF HEMT cooled preamplifier, which is referred to the JPA input. Thus, this calibration includes losses between the JPA and the HEMT in the measured noise value. These values are fully consistent with the specifications and with previous measurements obtained using the same amplifier.}

In order to find the added noise of an individual amplifier, we employ a spectrum analyzer to measure both gain $G$ and the improvement in signal-to-noise ratio $\Delta{}$SNR when the amplifier is turned on. The added noise of the amplifier $T_{\textrm{N}}$ can then be calculated from equation \cite{caves2012quantum} 

\begin{equation}
 T_{\textrm{N}}= T_\textrm{S}\left(\frac{1}{\Delta\textrm{SNR}}-\frac{1}{G}\right) - T_\textrm{Q}\left(1-\frac{1}{\Delta\textrm{SNR}}\right),
 \label{Eq:Addednoise}
\end{equation}
where $T_{\textrm{S}}$ is the calibrated HEMT system noise temperature before the addition of the amplifier. The second term on the right-hand side accounts for the quantum fluctuations already present at the amplifier input and becomes significant only when the system noise temperature approaches the standard quantum limit.

It should be noted that the standard quantum limit for added noise and system noise temperature are not the same. The SQL for added noise is half that of the system noise, as the added noise of the amplifier does not include the unavoidable quantum contribution of $\hbar\omega_\textrm{s}/2$ from the signal source. To circumvent this limit, one would have to squeeze the vacuum at the input of the amplifier at the signal frequency \cite{castellanos2008amplification,aasi2013enhanced}.

\subsection*{Results: Gain and noise performance}

For direct probing of noise in phase-sensitive parametric amplification, we constructed a measurement scheme where the amplified 
signal from the JPA is further amplified by a TWPA \cite{macklin2015near,esposito2022observation,qiu2023broadband} manufactured by VTT \cite{Perelshtein2021}. This allows us to estimate the noise performance of the JPA even at low gains. We use an Anapico APMS12G-4 signal generator to generate both the signal and the pump tones. The generated signals have relative phase stability of 3 mrad over 5 hours around the measurement frequency, which ensures phase stability between the two tones during the experiment.

At the beginning, the calibration for system noise temperature is obtained with both amplifiers off, as our calibration method requires heating up of the environment, which may affect JPA and TWPA performance. Added noise for each amplifier is then calculated from Eq. \ref{Eq:Addednoise}, after which the new system noise temperature is recalculated from Eq. \ref{Eq:Systemnoise}. Of note is that in the Friis formula, the quantum contribution to noise cannot be suppressed by the amplifier gain.

After the initial calibration, only the TWPA is biased in phase-preserving mode to provide $\sim 20$ dB with added noise of $T_\textrm{N,TWPA}\approx 700$ mK. At this point, the system noise temperature is $T_\textrm{S,TWPA} \approx 900$ mK. Table \ref{Table:Device_params} provides information about the operation parameters and performance of both devices. The intrinsic cavity loss rate of the JPA was determined from a standard $S_{21}$ measurement using complex circle fitting, yielding $\gamma=0.22\kappa$.

\begin{figure}[h]
    \noindent\centering{
    \includegraphics[width=1\linewidth]{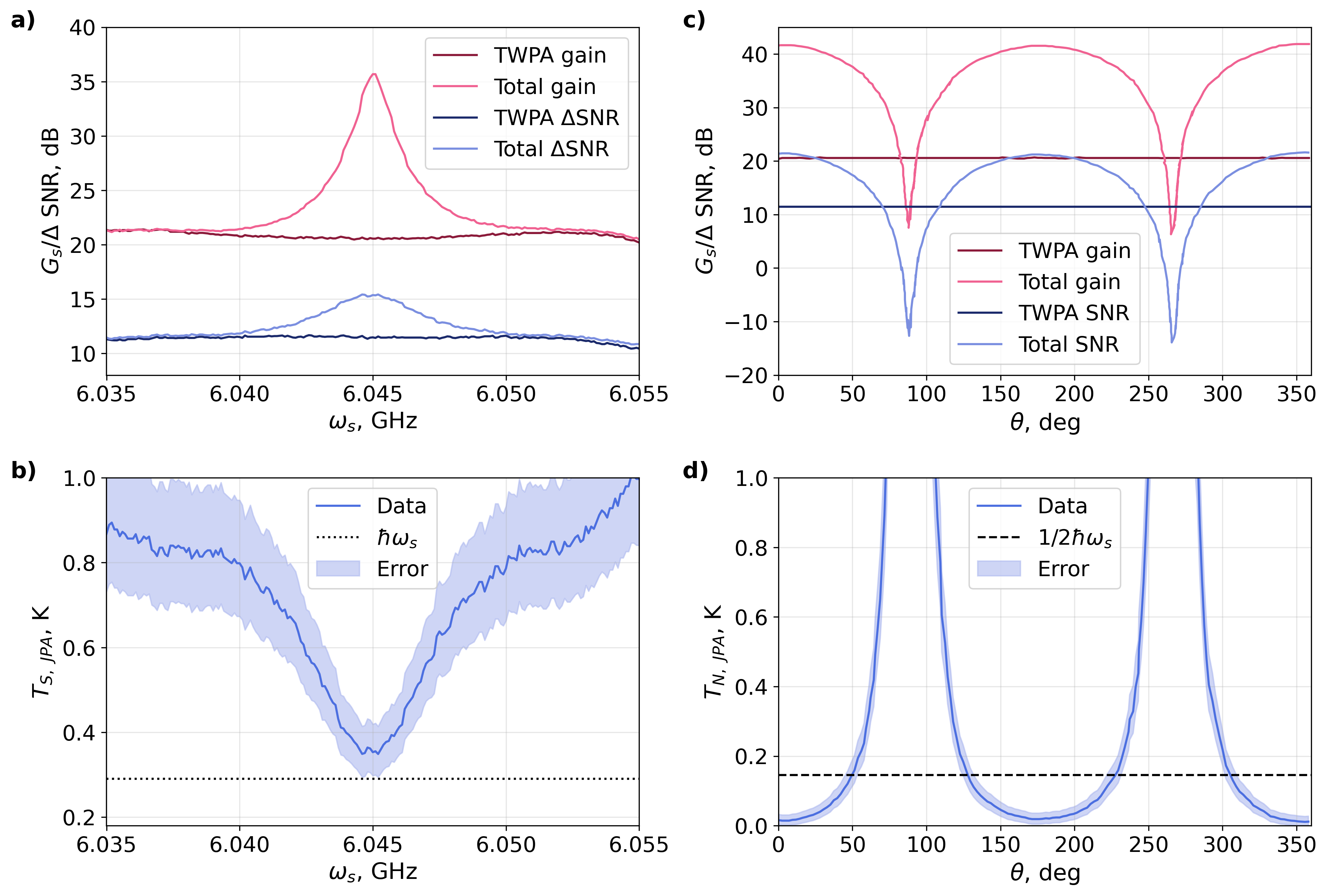}
    }
    \caption{
    \textbf{a)} Phase-preserving parametric amplification as a function of signal frequency, showing the narrow-band performance of the JPA on top of the TWPA performance. \textbf{b)} The initial system noise temperature (at $G_{\rm{JPA}}=0$ dB) of approximately $T_{\rm{S,TWPA}} \simeq 0.75$\;K is achieved in part due to the pre-amplification provided by the TWPA. (Right side) \textbf{c)} Gain as a function of relative phase difference between signal and pump $\theta$ in the degenerate operating mode of the JPA. \textbf{d)} Phase dependence of $T_N$ in degenerate operation. When the signal and pump are out of phase, deamplification occurs, and the input-referred noise appears to increase as a result.}
    \label{CombinedPPAndPSOp}
\end{figure}

For comparison, we include performance results on both phase-preserving and phase-sensitive parametric amplification. The phase-preserving operation of the JPA is shown on the left side of Fig. \ref{CombinedPPAndPSOp} and illustrates the low bandwidth nature of the JPA as compared to the TWPA and how the cascade is close to quantum-limited in noise performance, yielding a noise temperature of $T_{\textrm{S,JPA}} = 351 \pm 53$ mK. Here, the subscript JPA now indicates that the system noise temperature is being referred to the input of the JPA. In the phase-sensitive amplification mode, shown on the right side of Fig. \ref{CombinedPPAndPSOp}, the phase of the signal is varied with respect to the phase of the pump tone. When the local oscillator is phase-locked to the amplified quadrature, the signal and idler amplitudes add coherently, resulting in a 6 dB increase in the measured gain of the phase-sensitive JPA (See, e.g., Ref. \onlinecite{malnou2024low}).

\begin{table}[ht]
\centering
\begin{tabular}{ |p{3cm}|p{3cm}|p{3cm}|  }
\hline
\multicolumn{3}{|c|}{Device parameters} \\
\hline
 & JPA & TWPA \\
\hline
$\omega_p/2\pi$ & 6.045 GHz & 5.5 GHz \\
GBW & 400 MHz & 400 GHz \\
$P_{\textrm{Sat}}$ & $-120$ dBm & $-101$ dBm \\
$P_{\textrm{Pump}}$ & $-90$ dBm & $-66$ dBm \\
\hhline{~--}
$T_{\textrm{0}}$ & \multicolumn{2}{c|}{15 mK} \\
\hline
\end{tabular}
\caption{Summary of the measured device parameters, including the half-pump center frequency ($\omega_p/2\pi$), gain bandwidth product (GBW), saturation power at $\simeq$20~dB gain ($P_{\textrm{Sat}}$), pump power ($P_{\textrm{Pump}}$), and operating temperature ($T_{\textrm{0}}$).}
\label{Table:Device_params}
\end{table}

\begin{figure}[h]
    \noindent\centering{
    \includegraphics[width=1\linewidth]{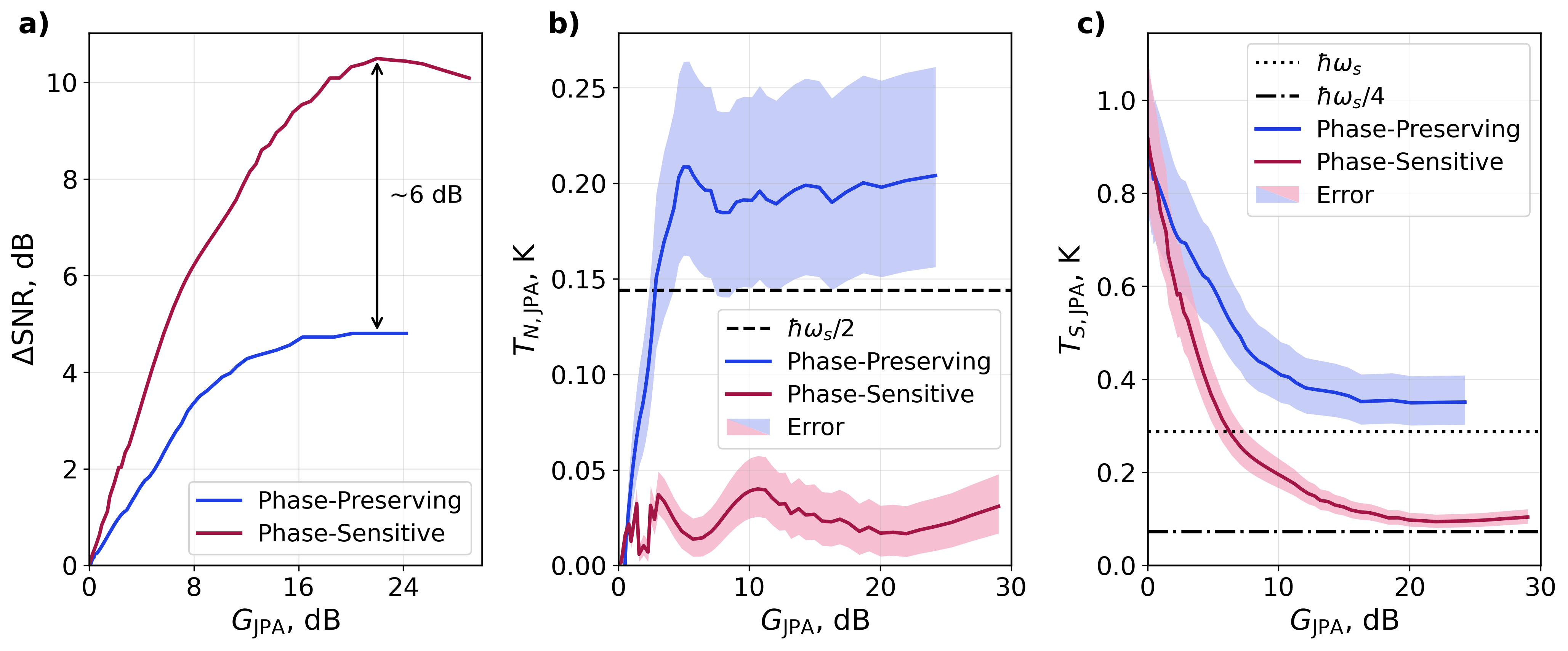}
    }
    \caption{\textbf{a)} Improvement in Signal-to-Noise ratio $\Delta\textrm{SNR}$ as a function of gain. Note tha approximately 6 dB enhancement in signal visibility is achieved in phase-sensitive operation. \textbf{b)} $T_{\textrm{N,JPA}}$ and \textbf{c)} $T_{\textrm{S,JPA}}$ as functions of $G_{\textrm{JPA}}$. At very low gain, no noise is added to the measurement by the JPA, while at sufficiently large gains the added noise becomes nearly constant. In contrast, the system noise temperature keeps lowering as the JPA gain actively suppresses noise contributions from following amplifiers in the cascade.
    }
    \label{GainAndDSNRBothRegimes}
\end{figure}

Figure~\ref{GainAndDSNRBothRegimes} shows the JPA added noise temperature, $T_{\mathrm{N,JPA}}$, and the corresponding system noise temperature, $T_{\mathrm{S,JPA}}$, as functions of JPA gain. In the phase-preserving regime, the added noise increases with gain and saturates at $200\pm54$ mK, corresponding to approximately $56$ mK above the standard quantum limit of $T_q=\hbar\omega_{\mathrm{s}}/(2k_{\mathrm{B}})=144$ mK. In contrast, during phase-sensitive operation the added noise saturates at only $17\pm12$ mK, well below the quantum limit for phase-preserving amplification. Expressed in terms of added input-referred noise, the phase-preserving mode adds $0.71$ photons, whereas the phase-sensitive mode adds only $0.06$ photons.

The corresponding system noise temperatures are shown in panel c) of Fig.~\ref{GainAndDSNRBothRegimes}. At sufficiently high JPA gain, the system noise is dominated by the JPA itself, including the unavoidable quantum fluctuations. In the phase-sensitive regime, the system noise temperature saturates at $94\pm12$ mK. Of this, approximately $89$ mK arises from the quantum fluctuations and the intrinsic added noise of the JPA, while the remaining $\sim5$ mK is contributed by the subsequent amplifier chain. Achieving such a low system noise temperature would not have been possible without the succeeding TWPA, which suppresses the noise contribution of the following amplifiers and enables the JPA to operate close to its fundamental limit.

Compared with previous demonstrations of phase-sensitive amplification, B. Yurke \cite{yurke_observation_1988} reported an added noise of 0.31 photons, L. Zhong \cite{zhong2013squeezing} reported 0.12 added photons, and D.J. Parker \cite{DPa_kipa_Grimsmo2022} reported 0.1 added noise photons above the quantum limit. The latter value, however, includes the noise contribution from the subsequent amplifier chain, making a direct comparison with our input-referred JPA added noise less straightforward. The main advantage of this work lies in the low-noise preamplification provided by the TWPA, which substantially reduces the noise contribution of the subsequent amplifier chain. In addition, the JPA offers a flux-tunable operating range from 5.4 to 6.1 GHz, enabling low-noise amplification across a broad frequency band. Finally, although the TWPA alone already provides near-quantum-limited performance in the phase-preserving regime, the additional gain and phase-sensitive amplification provided by the JPA are essential for reducing the overall system noise below the standard quantum limit. These results demonstrate the advantage of the JPA-TWPA cascade for ultra-low-noise microwave measurements.

\subsection*{Conclusion}

Utilizing a cascade of JPA\&TWPA amplifiers, we managed to reach an added noise temperature $T_{\textrm{N,JPA}} \simeq 17\pm 12$\;mK and a system noise temperature of $T_{\textrm{S,JPA}} \simeq 94\pm 12$\;mK in the phase-sensitive operating regime of the JPA at 6 GHz. This opens up the possibility to directly probe the state of a quantum device located before the JPA in the component chain, without having to subtract noise from the measurement \cite{mallet2011quantum}. Using a squeezed vacuum state for measurement preparation, JPA-based experiments can be performed below the quantum limit \cite{Yurke1984,aasi2013enhanced,Lehnert2019}. In addition, an ultimate-sensitivity amplifier chain could enhance the performance of various quantum communication protocols, including quantum secure direct communication \cite{shapiro2014secure,pan2025simultaneous} and continuous-variable quantum key distribution \cite{fesquet2024demonstration,liao2025experimental}. Degenerate 3WM amplifier operation can also be achieved using short traveling wave parametric amplifier sections with the benefit of flexible tunability of the employed frequency, which in transmission-line-cavity JPAs is approximately limited to the half-width of the resonance curve \cite{Lilja2026}.

\subsection*{Acknowledgements}

We acknowledge the team at VTT Technical Research Centre of Finland Ltd. for fabricating the devices: Joonas Govenius, Leif Grönberg, Robab Najafi Jabdaraghi, Janne Lehtinen, and Mika Prunnila. This work was supported by the Research Council of Finland projects 312295, 352926, and 374170 (CoE QMAT). The research leading to these results has received funding from the European Union’s Horizon 2020 Research and Innovation Programme, under Grant Agreement No.~824109 (EMP). The work of KP was supported by the European Union under Horizon Europe 2021-2027 Framework Programme, project MiSS (Microwave Squeezing with Superconducting (meta)materials) grant agreement ID: 101135868. IL is grateful for Vaisala Foundation of the Finnish Academy of Arts and Letters for a stipend. EM acknowledges a PhD scholarship from InstituteQ. The support of the Jane and Aatos Erkko Foundation (Future Makers SELQIT project) and the Keele Foundation (SuperC project) is also gratefully acknowledged.




\bibliography{Bibliography}

\end{document}